\documentclass{sig-alternate}   % For Latex2e

\usepackage[paper=letterpaper,width=7in,height=9.25in,centering]{geometry}
\usepackage{float}
\usepackage{amssymb,amsmath}
\usepackage{theorem}
\usepackage{graphics}
\usepackage{amsfonts}
\usepackage{mathrsfs}
\usepackage{amscd}
\usepackage{caption}
\usepackage{supertabular}
\usepackage{longtable}
\usepackage{color}
\usepackage{url}
\usepackage[plainpages=false,pdfpagelabels=true,colorlinks=true,citecolor=blue,hypertexnames=false]{hyperref}

\newtheorem{theorem}{Theorem}[section]

\newtheorem{proposition}[theorem]{Proposition}

\newtheorem{define}[theorem]{Definition}
\newtheorem{example}[theorem]{Example}

\newtheorem{corollary}[theorem]{Corollary}
\usepackage{tikz}

\tikzstyle{level 1}=[level distance=2cm, sibling distance=6.5cm]
\tikzstyle{level 2}=[level distance=3cm, sibling distance=3.5cm]

\tikzstyle{bag} = [text width=8em, text centered, font=\footnotesize]

\usetikzlibrary{arrows, shapes, trees}

\newcommand{\ignore}[1]{}

\floatstyle{ruled}
\newfloat{algorithm}{!h}{loa}
\floatname{algorithm}{Algorithm}

\def\qed{\hfil {\vrule height5pt width2pt depth2pt}}

\def\qed{\hfil {\vrule height5pt width2pt depth2pt}}

\def\u{U}

\def\x{X}

\def\CEssential{{\rm CheckEssential}}
\def\MCGBMain{{\rm MCGBMain}}
\def\CEssentialInBranch{{\rm CheckInBranch}}

\def\UpdateCGS{{\rm UpdateCGS}}
\def\MCGBSimpl{{\rm MCGBSimpl}}

\def\UpdateCGS{{\rm UpdateCGS}}

\def\Prep{{\rm Preprocess}}

\newcommand{\comment}[1]{}
\newcommand{\gr}{Gr\"obner\,}

\begin{document}

\conferenceinfo{ISSAC 2015,}{6-9 July, Bath, UK}

\title{An Algorithm for Computing a Minimal Comprehensive
  Gr\"obner\, Basis of a Parametric Polynomial
  System\titlenote{A preliminary extended abstract of this paper appeared in
 Proc. {\em Conference
    Encuentros de Algebra Comptacional y Aplicaciones (EACA),}
  Barcelona, Spain, June 2014, 21-25 \cite{KapurYangEACA14}.
 http://www.ub.edu/eaca2014/\\2014\_EACA\_Conference\_Proceedings.pdf. }\\}

\numberofauthors{2}

\author{
\alignauthor 
Deepak Kapur \\
       \affaddr{Department of Computer Science  }\\
       \affaddr{ University of New Mexico}\\
       \affaddr{Albuquerque, NM, USA}\\
       \email{kapur@cs.unm.edu }
\alignauthor Yiming Yang\\
	\affaddr{Department of Computer Science}\\
	\affaddr{ University of New Mexico}\\
	\affaddr{Albuquerque, NM, USA}\\
	\email{yiming@cs.unm.edu }
}
\date{\today}

\maketitle

\begin{abstract}
  An algorithm to generate a minimal comprehensive Gr\"obner\, basis of a
  parametric polynomial system from an arbitrary faithful
  comprehensive Gr\"obner\, system is presented. A basis of a parametric
  polynomial ideal is a comprehensive Gr\"obner\, basis if and only if
  for every specialization of parameters in a given field, the
  specialization of the basis is a Gr\"obner\, basis of the associated
  specialized polynomial ideal. The key idea used in ensuring
  minimality is that of a polynomial being essential with respect
  to a comprehensive Gr\"obner\, basis. The essentiality check is
  performed by determining whether a polynomial can be covered
  for various specializations by other polynomials in the
  associated branches in a comprehensive Gr\"obner\, system. The
  algorithm has been implemented and successfully tried on many
  examples from the literature.

\end{abstract}


\maketitle

\section{Introduction}

The concepts of a comprehensive \gr system (CGS) and a
comprehensive \gr basis (CGB) were introduced by
Weispfenning \cite{Weis1} to associate \gr basis like
objects for parametric polynomial systems (see also the notion of
a related concept of a parametric \gr basis (PGB) independently
introduced by Kapur \cite{Kapur}).  For a specialization of
parameters, a \gr basis of the specialized ideal can be
immediately recovered from a branch of the associated
CGS. Similarly, given a
CGB, one merely has to specialize it to construct a \gr
basis of the specialized ideal.

The above stated properties of CGS and CGB make them attractive in applications
where a family of related problems can be parameterized and
specified using a parametric polynomial system. For various
specializations, they can be solved by specializing a parametric
solution without having to repeat the computations.
Because of their applications, these concepts and related
algorithms have been well
investigated by researchers and a number of algorithms have been
proposed to construct such objects for parametric polynomial
systems (\cite{buildtree}, \cite{Weis2},\cite{SS1}, \cite{SS2},
\cite{SS3}, \cite{dispgb}, \cite{Nabeshima}, \cite{wibmer},
\cite{mccgs}, \cite{gbcover}, \cite{KSW1}, \cite{KSW2}).  An
algorithm that simultaneously computes a comprehensive
\gr system and a comprehensive \gr basis by
Kapur, Sun and Wang (KSW) \cite{KSW2} is particularly noteworthy
because of its many nice properties: (i) fewer segments
(branches) in a comprehensive \gr system generated by the
algorithm, (ii) all polynomials in a CGS and CGB are faithful meaning
that they are in the input ideal, and more
importantly, (iii) the algorithm has been found efficient in
practice \cite{Montes}.

For the non-parametric case, \gr bases have a very nice
property: once an admissible term ordering is fixed, every ideal
has a canonical \gr basis associated with it; a canonical
\gr basis is not only unique but also reduced and
minimal. This property is quite useful in many applications since
equality of two ideals can be easily checked by checking the
equality of their unique reduced minimal \gr bases. The
final goal of this research project is to work towards a similar property for
parametric ideals. The problem addressed here is to define a minimal
CGB associated with a parametric ideal once an admissible term
ordering (both on the parameters as well as variables) is fixed;
this seems to be the first step toward defining a canonical CGB.

There are some proposals in the literature for defining a
canonical CGB which are not satisfactory. Consider for instance,
Weispfenning's proposal in \cite{Weis2}: without reproducing his
definition, we give an example of a parametric ideal from his
paper generated by a basis $\{ f = u y + x, g = v z + x + 1 \}$;
Weispfenning reported $\{ f, g, h, -h \}$ as a canonical CGB of
the ideal, where $h$ is $v z - u y + 1$ using the lexicographic
ordering $z > y > x \gg v > u$.

We claim that each of $f, g, h$ is \textbf{essential} whereas
$-h$ is not: for any specialization $\sigma$, if $\sigma(h)$ is
in a GB of $\sigma(I)$, then $\sigma(-h)$ is reducible using
$\sigma(h)$ and vice versa; so only the \emph{smaller} of the two
has to be in a minimal CGB, which is $h$ because though their
leading coefficients only differ on the sign, $h$ is monic while
$-h$ is not. Obviously both $f$ and $g$ are essential: for
$\sigma$ in which $uv \neq 0$, the leading term of $\sigma(f)$ is $y$
whereas the leading term of the other two is $z$; for $\sigma$ in which $v = 0$
but $u \neq 0$, the leading term of $\sigma(g)$ is $x$ while the
leading term of the other two is $y$. Further, since $h = g - f$, the GB theory in
the nonparametric case would suggest that $h$ is not
essential, but that is not true: for the case $u = 0, v = 0$,
$\sigma(I) = \langle 1 \rangle$, however, $\{\sigma(f) = x,
\sigma(g) = x + 1\}$ is not a GB of $\sigma(I)$,
which implies that $h$ is essential. The algorithm proposed in the paper
generates $\{f , g, h\}$ as the minimal CGB of this parametric
ideal, regardless of the order in which these polynomials are
checked for essentiality (see Section 3).

Notice from this example that a minimal
CGB is not reduced in the conventional sense (since $h$ can be
reduced using $g$); further some of its specializations
are neither reduced nor minimal, whereas other specialization
(e.g. for $\sigma: u = 0, v = 0$), $\sigma(h)$
is both reduced and minimal.
There are examples also of minimal CGBs which are reduced but their
specializations are neither reduced nor minimal.

Montes and Wibmer \cite{gbcover} also define an object related to a canonical
CGS in \cite{gbcover}; but unfortunately, this object cannot be
used to generate a canonical CGB since elements in the CGS are not
faithful, i.e., not in the original ideal.

In this paper, we give an algorithm for generating a minimal CGB
from a given CGS having the property that associated with every
segment describing parameter specialization, there is a GB for
that specialization in which every polynomial is in the original
ideal, i.e.\ faithful.  To illustrate the proposed algorithm, we
assume that this algorithm takes as input,
the CGB generated by the KSW 
algorithm \cite{KSW2},  which is the union of all \gr
basis associated with each branch in the associated CGS generated by the KSW
algorithm. Recall that every polynomial in the CGB
generated by the KSW algorithm is faithful since every polynomial
in the GB of every segment of its CGS is faithful. In a related
paper \cite{Completion}, we have proposed a Buchberger like
completion algorithm for computing a minimal CGB directly from a
basis of a parametric ideal.

Given a CGB and the associated CGS of a parametric ideal as well as an admissible term
ordering as input, the proposed algorithm 
has to deal with two key issues: (i) \emph{redundnacy} in a CGB:
since the CGB is constructed as the union of different branches
of a CGS, it may include polynomials which while necessary and
minimal for a subset of specializations, can be covered by other
polynomials needed for a different subset of specializations,
(ii) nonredundant polynomials could be further simplified using
other polynomials. It 
first checks whether every polynomial in CGB is \emph{essential} in the sense
that there is at least one specialization of parameters under
which it is essential (necessary) to include the specialization
of the polynomial in a \gr basis of the associated
specialized ideal. 
This check is performed by testing whether the polynomial can be
covered by other polynomials in CGB; this is done by identifying
every branch in the associated CGS in which the polynomial
appears in the associated \gr basis and finding whether
other polynomials in the CGB can cover it for all specializations
corresponding to the branch. If a polynomial is discovered to be
not essential, it is replaced  by other polynomials covering it in branches
in CGS it used to appear, and also discarded from the CGB.
The subset of essential polynomials in a given 
CGB is a minimal CGB. 

As will be shown later, there can be multiple minimal CGBs being generated from a
given CGS even when an admissble term ordering is fixed. To obtain the least minimal
CGB among them, \emph{simplification} (similar to normalization) on an essential polynomial by other essential
polynomials is attempted; if the result is different from the
original polynomial and covers the original polynomial with the
help of other essential polynomials in 
the CGB, it can replace the
original polynomial in the CGB. Despite this simplification, it
is still possible to change heuristics in the proposed
algorithm to generate different minimal CGBs.

After discussing preliminaries in the next section, the algorithm
is presented in Section 3. The concept of an essential polynomial
in a CGB is defined, and the essentiality check is
explained. Covering of a polynomial under a set of
specializations defined by a segment by other polynomials is
defined. The termination and correctness of the algorithm are
sketched. It is shown that the proposed algorithm can generate
multiple minimal CGBs depending upon the order in which
essentiality check is performed. It is proved that if the
essentiality check is performed in desecnding order of
polynomials, then the resulting minimal CGB is the least among
all CGBs which are subsets of the input CGB. To generate minimal
CGBs which are not subsets of the input CGB, 
the concept of simplification of an essential polynomial by other
essential polynomials is introduced. Section 4 discusses the
performance of our implementation of the proposed algorithm. It
is shown that most CGBs computed by various algorithms have
redundnant polynomials; in particular, the KSW algorithm can
sometimes generate CGBs in which about half of the polynomials are redundant.

\section{Preliminaries}

The reader may consult \cite{Cox} for definitions. Below we
introduce notation and definitions necessary for the paper. 
Let $K$ be a field, $L$ an algebraically closed field extension
of $K$, $U$ and $X$ the sets of parameters and variables
respectively.

Let $>$ be an admissible total term ordering in which $X \gg U$,
i.e.\ variables are bigger than parameters. In the ring $K[X,
U]$, which regards parameters as the same as variables, for a
polynomial $f \in K[X, U]$, $LC_U(f)$ and $LT_X(f)$ are defined
as its leading parametric coefficient and leading term w.r.t.\
$>$ respectively. For example, let $f = 32(u-1)x^2 + 4uy$, where
$U = \{u\}$, $X = \{x, y\}$, and $>$ is a lexicographic term
ordering with $x > y \gg u$. Then $LC_U(f) = 32(u-1)$ and
$LT_X(f) = x^2$. 
%Then for a polynomial $g$ in the ring $L[X]$,
%its leading coefficient and leading term w.r.t.\ $>$ are denoted
%as $LC(g)$ and $LT(g)$ respectively. Let $g = 47x^2y^2 + x3$,
%with $>$ being a graded lexicographic order with $x > y$, then
%$LC(g) = 47$ and $LT(g) = x^2y^2$.

A specialization $\sigma$ is a ring homomorphism from $K[U]$ to
$L$, which can be extended canonically to $K[U][X] \rightarrow
L[X]$ by keeping the identity on variables. For a polynomial $f
\in K[U][X]$, $\sigma$ is given by $f \rightarrow f(v_1, v_2,
\dots, v_m)$, where $v_1, \dots, v_m \in L$. In brief, denote
this image of $f$ as $\sigma_{\overline{v}}(f)$, where
$\overline{v} = (v_1, \dots, v_m) \in L^m$, or simply $\sigma(f)$
if it is clear from the context. 

\vspace{-3mm}
\begin{define}
\label{Def Segment}
Let $E, N$ be subsets of $K[U]$, then the tuple $(E, N)$ is called a \textbf{parametric segment}. 
An associated \textbf{constructible set} $A$ is given by
$\mathbb{V}(E) - \mathbb{V}(N)$, where $\mathbb{V}(E)$ is the algebraic variety
(the zero set) of $E$ in $L$. $(E, N)$ is consistent if $A
\neq \emptyset$.
\end{define}

\vspace{-5mm}
\begin{define}
\label{Def CGS}
Given an ideal $I = \langle F \rangle \subseteq K[U][X]$, where
$F$ is finite, and an admissible term order $>$, let $A_1, \dots,
A_l$ be constructible sets of $L^m$, and $G_1, \dots, G_l$
subsets of $K[U][X]$, and $S$ a subset of $L^m$ such that $S
\subseteq A_1 \cup \cdots \cup A_l$. Then a \textbf{comprehensive
  \gr system (CGS)} of $I$ on $S$ w.r.t.\ $>$ is a finite set $CGS = \{(A_1, G_1), \dots, (A_l, G_l)\}$, where for $\forall 1 \leq i \leq l$, $\sigma_i(G_i)$ is a \gr basis of the ideal $\sigma_i(I)$ on $L[X]$ under $\forall \sigma_i \in A_i$.
\par
Each $(A_i, G_i)$ is called a branch of $CGS$. Specifically, if
$S = L^m$, then $CGS$ is called a comprehensive \gr system (CGS) of $I$.
\end{define}

\vspace{-1mm}
The above definition of a CGS does not require that $G_i$ be a subset of $I$; in fact, there are
algorithms (\cite{buildtree}, \cite{mccgs}, \cite{SS3}, \cite{KSW1}) for computing a CGS of an $I$ in which $G_i$ need not
be a subset. A CGS is called \emph{faithful} if and only if each
$G_i \subseteq I$. 

\vspace{-3mm}
\begin{define}
\label{Def CGB}
Given an ideal $I \subseteq K[U][X]$, $S \subseteq L^m$ the
parameter space, and an admissible term order $>$, let $\mathcal{G}$ be a
finite subset of $K[U][X]$. $\mathcal{G}$ is called a
\textbf{comprehensive \gr basis (CGB)} of $I$ on $S$ w.r.t.\ $>$,
if for $\forall \sigma \in S$, $\sigma(\mathcal{G})$ is always a \gr basis
of the ideal $\sigma(I)$ on $L[X]$. Specifically, if $S = L^m$,
$\mathcal{G}$ is a CGB of $I$. 
\end{define}

\vspace{-1mm}
Note that the above definition of a CGB requires it to be
faithful. From a faithful CGS, it is easy to compute the
associated CGB by taking the union of the set of  polynomials in
each branch of the CGS.
Further, a CGB  can be defined on
a constructible set (or equivalently, a parametric segment). The following proposition holds for any $CGS$ generated by the KSW algorithm in \cite{KSW2}

\vspace{-3mm}
\begin{proposition}
\label{Prop CGB on S}
If a $CGS$ of an ideal $I \subseteq K[U, X]$ w.r.t.\ an admissible term order $>$ is faithful, for every branch ($A_i$, $G_i$) $\in$ $CGS$, $G_i$ is a CGB of $I$ on $A_i$ w.r.t.\ $>$.
\end{proposition}

\vspace{-1mm}
Further, a special kind of CGB is defined as follows:

\vspace{-3mm}
\begin{define}
\label{Def MCGB}
Given a CGB $\mathcal{G}$ of an ideal $I \subseteq K[U][X]$ w.r.t.\ an admissible term order $>$, $\mathcal{G}$ is \textbf{minimal} if the following conditions are true --
\begin{itemize}
	\item[(i)]
	No proper subset of $\mathcal{G}$ is a CGB of $I$ w.r.t.\ $>$;
	\item[(ii)]
	For $\forall g \in \mathcal{G}$, $LC_U(g)$ is a monic polynomial in $K[U]$.
\end{itemize}
\end{define}

\vspace{-1mm}
The $CGS$ and the associated CGB $\mathcal{G}$ computed by the KSW
algorithm is adapted to Definition \ref{Def MCGB}, by making
polynomials in each branch of $CGS$ and every polynomial
in $\mathcal{G}$ have their leading coefficients as monic
polynomials in $K[U]$.

\section{Algorithm For Generating Minimal CGB}

The proposed algorithm takes a CGB $\mathcal{G}$ and its associated faithful CGS as input, and outputs a minimal CGB (MCGB) of the same ideal. The structures generated by the KSW algorithm \cite{KSW2} satisfies this requirement. Two key operations are performed on $\mathcal{G}$ to get an MCGB: (i) removal of non-essential (or redundant) polynomials in $\mathcal{G}$, and (ii) simplification of essential polynomials by
other polynomials in $\mathcal{G}$. In subsequent sections, we discuss
in detail how these checks are performed. We start with a top
level description of the algorithm first.

A minimal \gr basis of an ideal
$I \subseteq K[X]$ is achieved by removing unnecessary
polynomials from an arbitrary \gr basis of $I$. 
Analogously, the concept of an \emph{essential} polynomial in a CGB is defined:

\vspace{-3mm}
\begin{define}
\label{Def Essential Poly}
Given a CGB $\mathcal{G}$ of some ideal $I \subseteq K[U][X]$, a polynomial $p \in \mathcal{G}$ is called \emph{essential} w.r.t.\ $\mathcal{G}$ iff $\mathcal{G} - \{p\}$ is not a CGB of $I$ anymore. Otherwise, $p$ is \emph{non-essential} iff $\mathcal{G} - \{p\}$ remains to be a CGB of $I$.
\end{define}

\vspace{-1mm}
We are abusing the notation somewhat since in checking whether
$p \in \mathcal{G}$ is essential, only polynomials in
$\mathcal{G} - \{ p \}$ are considered.

\vspace{-1mm}
The following proposition intuitively states that a polynomial
$p$ in the above $\mathcal{G}$ is essential 
if under at least one
specialization $\sigma$, $\sigma(p)$ is necessary for
$\sigma(\mathcal{G})$ to be a \gr basis of the specialized ideal
$\sigma(I)$. Further,

\vspace{-2mm}
\begin{proposition}
\label{Prop Essential Specialization}
Given a CGB $\mathcal{G}$ of some ideal $I$, a polynomial $p \in \mathcal{G}$ is essential w.r.t.\ $\mathcal{G}$ if and only if there exists a specialization $\sigma$ such that for $\forall q \in \mathcal{G} - \{p\}$, $LT(\sigma(q))$ cannot divide $LT(\sigma(p))$.
\end{proposition}

\vspace{-3mm}
\begin{corollary}
\label{Corollary Essential Still Essential}
Given a CGB $\mathcal{G}$ of an ideal $I$ and a polynomial $p \in
\mathcal{G}$ that is essential w.r.t.\ $\mathcal{G}$, $p$ remains essential w.r.t.\ any CGB of $I$ which is also a subset of $\mathcal{G}$.
\end{corollary}

%\vspace{-1mm}
%Once a polynomial in $\mathcal{G}$ is checked being essential, there is no need to recheck it anymore in the computation.

\vspace{-1mm}
A minimal CGB from a given CGB is computed by removing non-essential polynomials
from it.

\subsection{When is a multiple of a polynomial in CGB redundant?}
\label{Section Multiple}

An easy and obvious redundancy check is when a CGB contains a
polynomial as well as its multiple with a polynomial purely in
parameters as the multiplier. For any specialization, the
multiplier evaluates to a constant.
Since the KSW algorithm computes the RGB in each branch
w.r.t.\ $K[X, U]$, a polynomial and its multiple can be part of
the output in different branches as illustrated below.

\vspace{-3mm}
\begin{example}
\label{Example Parametric Multiple}
Given $I = \langle f \rangle = \langle (a^3-b^3)x^2 + (a^2+b^2+1)x + (a-b)(b+2) \rangle \subseteq K[a, b][x]$ and a lexicographic term order $>$ with $x \gg a > b$.
\par
A CGB of $I$ computed by the KSW algorithm is 
\vspace{-2mm}
\[
	\mathcal{G} = \{ f, \text{ } g = (b^3 - \frac{1}{4}b^2 + \frac{3}{2}b + \frac{1}{2}) f, \text{ }h = (a+b) f \}.
\]
\end{example}

\vspace{-2mm}
Both $g$ and $h$ are multiples of $f$ with the multipliers being
polynomials in $K[U]$ (i.e.\ polynomials only in parameters). 
It is easy to see that both $g$ and $h$ are redundant, and
after removing them, the resulting set $\mathcal{G}' = \{f\}$ is
still a CGB of $I$. In general,

\vspace{-3mm}
\begin{proposition}
\label{Prop Param Multiple}
Let $\mathcal{G}$ be a CGB of ideal $I \subseteq K[U][X]$ w.r.t.\ $>$. If there is $S = \{c_1 f, \dots, c_k f \} \subseteq \mathcal{G}$ with $f \in \mathcal{G}$ and $c_1, \dots, c_k \in K[U]$, then $\mathcal{G} - S$ is still a CGB of $I$.
\end{proposition}

\vspace{-1mm}
The CGB output of the KSW algorithm is thus 
preprocessed to remove such multiples of polynomials with
multipliers in $K[U]$ if they are present; 
The asssociated CGS is updated by replacing $c_i f$ by $f$ in its branches.
By this removal, we
achieve a CGB of the same ideal of a smaller size along with a
simpler CGS. %thus making it
%easier to analyze of the \gr basis of the given ideal $I$ under
%specialization.

\subsection{Key Ideas of the Algorithm}
\label{Section MCGB Algorithm}

The algorithm below is given
the CGB $\mathcal{G}$ and its associated $CGS$ of
$I$ computed by the KSW algorithm as input. It preprocesses $\mathcal{G}$ as discussed in Section
\ref{Section Multiple}; if any polynomial is deleted in this
step, the associated $CGS$ is updated accordingly.
Each polynomial in the result is checked for being essential.
\begin{enumerate}
\item $p$ is not essential: 
  $p$ is removed from $\mathcal{G}$, and $CGS$ is updated as in
  Section \ref{Section Update CGS}, replacing $p$ by other polynomials which
  cover $p$.

\item $p$ is essential: $p$ is kept in $\mathcal{G}$ without
  changing $CGS$.
\end{enumerate}
\newpage
\noindent\rule{8.5cm}{0.4pt}\\{\bf Algorithm $\MCGBMain(E, N, F)$}\\
{\bf Input: } $(E, N, F)$: $E$, $N$, finite subsets of $k[\u]$; $F$, a finite subset of $(K[\u,\x])^2$.\\
{\bf Output: } $\mathcal{M}$: A minimal comprehensive \gr basis of the ideal $\langle F \rangle$ w.r.t.\ the given term ordering $>$.

\begin{tabular}{l l}\\
1. & $(CGS, \mathcal{G}) := KSW(E, N, F)$;\\
2. & \textbf{if} $CGS = \emptyset$ \textbf{or} $\mathcal{G} = \emptyset$ \textbf{then return} $\emptyset$; \textbf{endif}\\
3. & $(CGS, \mathcal{G}) := \Prep(CGS, \mathcal{G})$;\\
4. & $\mathcal{G} := SortDesc(\mathcal{G})$;\\
5. & $\mathcal{M} := \mathcal{G}$;\\
6. & \textbf{for each} $p_i \in \mathcal{G}$ from $i := 1$ to $|\mathcal{G}|$:\\
7. & $\quad$ $L := \CEssential(p_i, \mathcal{M}, CGS)$;\\
8. & $\quad$ \textbf{if} $L \neq \emptyset$ \textbf{then}\\
9. & $\quad \quad$ $CGS := \UpdateCGS(CGS, p_i, L)$;\\
10.& $\quad \quad$ $\mathcal{M} := \mathcal{M} - \{ p_i \}$;\\
    & $\quad$ \textbf{endif}\\
    & \textbf{endfor}\\
11. & \textbf{return} $\mathcal{M}$;\\
\end{tabular}

\noindent\rule{8.5cm}{0.4pt}
\smallskip

In \textit{\MCGBMain}, procedures \textit{KSW} in Line $1$ and
\textit{\Prep} in Line $3$ represent the KSW algorithm and the
preprocessing respectively. As discussed later in Section
\ref{Essentiality Check}, the \textit{\CEssential} procedure in
Line $6$ returns an empty list if the input polynomial is found to be essential;
otherwise, it returns a list of branches with $p$ substituted by
its corresponding coverings. This list is used for updating $CGS$,
using the \textit{\UpdateCGS} procedure in Line $8$. 

The result of the algorithm is sensitive to the order in which
polynomials are checked for being essential.
The \textit{SortDesc} procedure in Line $4$ sorts
polynomials in $\mathcal{G}$ in a descending order w.r.t.\ $>$,
so that the essentiality check starts from the largest polynomial
to the least one. As proved later in Section \ref{Section
  Correctness}, 
polynomials checked in
decending order compute the smallest MCGB under the set ordering
w.r.t.\ $>$.

\subsection{An Illustrative Example}
Before discussing further details of the above algorithm, we
illustrate the key concepts of esssentiality and covering of a
polynomial by other polynomials using a simple example below.
\vspace{-3mm}
\begin{example}
\label{Example Illustrative}
Given $I = \langle (a-2b)x + y^2 + (a+b)z,\ a^2x+y+bz \rangle
\subseteq K[a, b][x, y, z]$ and a lexicographic term order such
that $x > y > z \gg a > b$. The KSW algorithm computes the following $CGS$:

\begin{center}
\begin{tabular}{|c|c|c|c|}\hline
\ & segment & basis & LT \\\hline
$1$ & $(\emptyset,\ \{ab\}\ )$ & $\{f_1, f_2\}$ & $\{y^2,\ x\}$ \\\hline
$2$ & $(\ \{b\},\ \{a\}\ )$ & $\{f_2, f_3\}$ & $\{y^2,\ x\}$ \\\hline
$3$ & $(\ \{a,\ b\},\ \{1\}\ )$ & $\{f_2\}$ & $\{y\}$ \\\hline
$4$ & $(\ \{a\},\ \{b\}\ )$ & $\{f_4, f_5\}$ & $\{y, x\}$ \\\hline
\end{tabular}
\end{center}
and the associated CGB 
\allowdisplaybreaks
\begin{align*}
	\mathcal{G} = \{ & f_1 = a^2y^2 + (-a+2b)y + \phi z,\\
	& f_2 = b^2x-\frac{a+2b}{4}y^2 + \frac{1}{4}y - \frac{\psi}{4}z,\\
	& f_3 = (a-2b)x + y^2 + (a+b)z,\\
	& f_4 = abx - \frac{a}{2}y^2 + \frac{1}{2}y - \frac{\theta}{2}z,\\
	& f_5 = abxy - \frac{a-2b}{2}x - \frac{a}{2}y^3 - \frac{a^2}{4}y^2z - \frac{2a^2+2ab-a}{4}yz\\
	& \quad \quad  - \frac{\phi}{4}z^2 - \frac{a+b}{2}z \},
\end{align*}
where $\phi = a^3+a^2b-ab+2b^2$, $\psi = a^2+3ab+2b^2-b$ and
$\theta = a^2+ab-b$.
\end{example}

\vspace{-2mm}
Preprocessing does not change
$\mathcal{G}$ since there are no polynomials multiple of each
other. Since $f_1 < f_2 < f_3 < f_4 < f_5$, the essentiality check starts from $f_5$ in the descending order. 

It suffices to check for specializations corresponding to branches where $f_5$
appears in the $CGS$ whether $f_5$ can be covered by other
polynomials in $\mathcal{G}$.  $f_5$ appears only in Branch 4
with the specializations $A_4: a = 0, b \neq 0$, and its leading term is
$x$. Since $G_4$ is minimal under $A_4$ due to the KSW algorithm,
it is enough to check whether polynomials in $\mathcal{G} -
G_4 = \{f_1, f_2, f_3\}$ can cover it for specializations of
$A_4$.  $f_2$ contains $x$ with coefficient $b^2 \neq 0$ and has
no higher term, so $\{f_2\}$ covers $f_5$ under $A_4$. Thus,
$f_5$ is not essential; it can be replaced by $f_2$ in the $CGS$
and deleted from $\mathcal{G}$ giving a smaller CGB of $I$. $G_4$
in $CGS$ becomes $\{f_4, f_2\}$ and $\mathcal{G} = \{f_1, f_2,
f_3, f_4\}$.

The essentiality of $f_4$ is checked similarly . It appears only
in Branch $4$ and its leading term is $y$ under $A_4$.  Further,
only $\{f_1, f_2\}$, which is $\mathcal{G} -G_4$ can possibly
cover it.  $f_1 $ contains $y$ with coefficient $-a+2b \neq 0$;
$f_1$ has a higher term $y^2$, but its coefficient $a^2 = 0$,
thus implying that the leading term of $f_1$ under $A_4$ is
determined to be $y$. So $f_1$ covers $f_4$ under $A_4$, implying
$f_4$ is non-essential. Update $\mathcal{G}$ and $CGS$ by
deleting $f_4$ and replacing $f_4$ by $f_1$ respectively:
$\mathcal{G} = \{ f_1, f_2, f_3 \}$, and $G_4$ becomes $\{ f_1,
f_2 \}$ in $CGS$.

$f_3$ appears only in Branch $2$ has $x$ as its leading
term $x$ in this branch. No polynomial in $\mathcal{G} - G_2$
contains $x$, implying $f_3$ has no covering under $A_2$. Thus $f_3$
is essential; consequently, $CGS$ and $\mathcal{G}$ do not change.

%%DK: if something appears in all branches, is it not essential then?
To check $f_2$ for essentiality,  all branches need to be considered. In Branch $1$,
the leading term of $f_2$ is $x$. $f_3$ is the only polynomial in
$\mathcal{G} - G_1$, which contains $x$ and has no higher
term. But the coefficient $a-2b$ is not determined. This means
$f_3$ covers $f_2$ only under the subsegment $A_{11} = A_1 \cup
\{-a+2b \neq 0\}$ which is a subset of $A_1$, but not under
$A_{10} = A_1 \cup \{-a+2b = 0\}$. There is no other polynomial
that has $x$ in it, so $f_2$ has no covering under $A_{10}$. Thus $f_2$ is
essential, and there is no need to consider other branches.

Finally, to check $f_1$ for essentiality, branches $1$ and $4$
need consideration. In
Branch $1$, the leading term of $f_1$ is $y^2$. $f_3$ is the only
polynomial in $\mathcal{G} - G_1$ and it contains $y^2$ with
coefficient $1 \neq 0$. But it has a higher term $x$ with
coefficient $a-2b$ that is not determined under $A_1$. So it cannot cover
$f_1$ under segment $A_{11} = A_1 \cup \{a-2b \neq 0\}$. Since
there is no other polynomial left to cover $f_1$, $f_1$ has no covering under
$A_{11}$. $f_1$ is essential, and the algorithm after having
found $f_4, f_5$ as nonessential with a smaller 
$CGB = \{ f_1, f_2, f_3\}$ consisting only of essential polynomials, with the updated smaller $CGS$ as well.

\vspace{-2mm}
\subsection{Essential Polynomial}
\label{Section Essential Poly}

\vspace{-2mm}
\subsubsection{Covering}
\label{Section Covering}

As illustrated by the above example, the essentiality check of a
polynomial $p$ in a CGB is performed by identifying other
polynomials in the CGB which can possibly \emph{cover} $p$.  For
a branch with an associated segment $A_i$ in a $CGS$ in which $p$
appears, only those polynomials in the CGB can cover $p$ that
have the leading term of $p$ w.r.t. $A_i$ appearing in them with
a possibly nonzero coefficient.  Since $p$ in the CGB may be
contributed by many branches in the CGS, for every such branch
and its associated segment, we need to check whether polynomials
appearing in other branches of CGS can cover $p$.  We exploit the
strucuture of the associated CGS computed by the KSW algorithm
and its properties to perform this check (particularly that each
branch $(A_j, G_j) \in CGS$, $G_j$ is a minimal \gr basis under
specializations in $A_j$ and all branches are disjoint). It is
thus not necessary to consider other polynomials in the same
branch as $p$ and further, we only need to consider polynomials
which have the leading term of $p$ under the segment under
consideration appearing with a possibly nonzero coefficient.  If
a single polynomial $q$ covers $p$ for the branch under
consideration, then $q$ can replace $p$ in that branch as well as
in the GB of the CGS of that branch.  For different branches in
which $p$ appears, there may be different such $q$'s covering
$p$. If for at least one branch in the CGS in which $p$ appears,
it cannot be covered, then $p$ is declared essential and kept in
the CGB as well as the CGS.  If $p$ can be covered in all
branches of the CGS in which it appears, then $p$ is not
essential and can thus be discarded from the CGB and the CGS;
further, $p$ is replaced by the corresponding $q$'s in respective
branches in the CGS.

It can be that for a particular branch in which $p$ appears, $p$
cannot be covered by a single polynomial but instead multiple
polynomials are needed for covering it. This especially arises if
the leading term of $p$ appears in another polynomial in CGB
but whose coefficient under the segment cannot be completely determined to be nonzero.
Then, the branch (i.e.,
segment of specializations) needs to
be split into multiple sub-branches so that multiple polynomials
can cover $p$. This aspect is also illustrated below in the next
subsection as well as in the case of the essentiality check of
$f_1, f_2$ in the example above.

\vspace{-3mm}
\begin{define}
\label{Def Covering General}
Given a CGB $\mathcal{G}$ of an ideal $I$, a finite set $Q = \{q_1, \dots, q_n\} \subseteq \mathcal{G} - \{p\}$ is said to be a \textbf{covering} of a polynomial $p \in \mathcal{G}$ over a set $A$ of specializations, if for $\forall \sigma \in A$, there is some $q_i \in Q$ such that $LT(\sigma(q_i)) \mid LT(\sigma(p))$.
\end{define}

\vspace{-3mm}
\begin{proposition}
\label{Prop Covering}
Given a CGB $\mathcal{G}$ and the associated $CGS$ of $I$ generated by
the KSW algorithm, $p \in \mathcal{G}$ is non-essential
w.r.t.\ $\mathcal{G}$, iff in each of branch $(A_i, G_i) \in CGS$
where $p$ appears (i.e.\ $p \in G_i$), $p$ has a covering $Q_i =
\{q_1, \dots, q_n\} \subseteq \mathcal{G} - G_i$, such that for
$\forall \sigma_i \in A_i$, there is some $q_j \in Q$ with
$LT(\sigma_i(q_j)) = LT(\sigma_i(p))$.

\end{proposition}

\subsubsection{Branch Partition}
\label{Section Branch Partition}

As alluded above,
a polynomial $p$ in a branch corresponding to the segment $A_j$
may not be covered by a single polynomial $q$ but
instead may require a set of polynomials $Q = \{q_1, \dots,
q_n\}$  with each polynomial in $Q$ only covers $p$ in a proper subset of $A_j$, but their union is $A_j$.

For example, in $K[u, v][y, x]$ and a lexicographic
term order with $y > x \gg u > v$, in a branch $A_j =
(\{u^2-v^2\},\ \{u\}) \in CGS$ of an ideal, $G_j = \{p =
(u^2-v^2)y + ux\}$ and $Q = \{q_1 = (u+v)x, \text{ } q_2 =
(u-v)x\}$. To check if $Q$ can cover $p$ in $A_j$, it is easy to see that both $q_1$ and $q_2$ partially cover $p$, since neither of their leading coefficients is determined. So $A_j$ is partitioned w.r.t.\ $u+v$:
$q_1$ covers $p$ in $A_{j1}$ and $q_2$ covers $p$ in
$A_{j0}$. So $Q = \{q_1, q_2\}$ is a covering of $p$ in
$A_j$. This kind of branch partition deals with the case when the
leading term of $p$ appears in $q$ but with the coefficient that
is not determined to be nonzero.

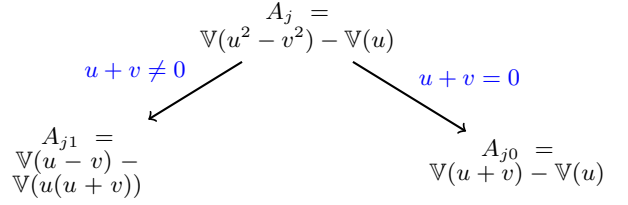
\begin{tikzpicture}[scale=.90, grow=down]
\node[bag] (root) {$A_j = $\\ $\mathbb{V}(u^2-v^2)-\mathbb{V}(u)$}
	child {
		node[bag] (s1) {$A_{j1} = $\\ $\mathbb{V}(u-v) - \mathbb{V}(u(u+v))$}
		edge from parent[draw=black, thick, ->] 
		node[above left] {$\textcolor{blue}{u+v \neq 0}$}
	}
	child {
		node[bag] {$A_{j0} =$\\$ \mathbb{V}(u+v)-\mathbb{V}(u)$}
		edge from parent[draw=black, thick, ->]
		node[above right] {$\textcolor{blue}{u+v = 0}$}
	};
\end{tikzpicture}

\subsection{Algorithm of Checking Essentiality}
\label{Essentiality Check}

%The essentiality check algorithm below is based on 
%the discussion in \ref{Section Essential Poly}.

\vspace{-2mm}
\noindent\rule{8.5cm}{0.4pt}\\{\bf Algorithm $\CEssential(p, \mathcal{G}, CGS)$}\\
\textbf{ Input: } $p$: a polynomial in $\mathcal{G}$ whose
essentiality is being checked; $\mathcal{G}$: a CGB of some ideal; $CGS$: the associated CGS of $\mathcal{G}$.\\
\textbf{ Output: } An empty list, if $p$ is essential w.r.t.\ $\mathcal{G}$; a non-empty list, otherwise.\\
\begin{tabular}{l l}\\
1. & $\mathscr{L} := \emptyset$;\\
2. & \textbf{for each} $B_i = (A_i, G_i) \in CGS$ with $p \in G_i$:\\
3. & $\quad$ $L_i := \CEssentialInBranch(p, \mathcal{G} - G_i, B_i)$;\\
4. & $\quad$ \textbf{if} $L_i = \emptyset$ \textbf{then} \textbf{return} $\emptyset$ \textbf{endif}\\
5. & $\quad$ $\mathscr{L} := \mathscr{L} \cup L_i$;\\
   & \textbf{endfor}\\
6. & \textbf{return} $\mathscr{L}$;\\
\end{tabular}\\
\noindent\rule{8.5cm}{0.4pt}

As illustrated above, to check whether a polynomial $p$ in a CGB
$\mathcal{G}$ is essential, the algorithm goes through each
branch $(A_i, G_i)$ where $p \in G_i$, looking for a covering of
$p$ by polynomials in $\mathcal{G} - G_i$.  In
\textit{\CEssentialInBranch}, let $(A_i, G_i)$ be a branch where
$p \in G_i$, and let $t$ be the leading term of $p$ under $A_i$;
then a set of candidate polynomials $G_{can}$ is first computed
from $\mathcal{G} - G_i$ by removing polynomials in which the
coefficient of the term $t$ is determined to be $0$. In Example
\ref{Example Illustrative}, to check $f_5$ under $A_4$ with $t =
x$, $G_{can} = \{f_2, f_3\}$ is computed out of $\mathcal{G} -
G_4 = \{f_1, f_2, f_3\}$, since $f_1$ contains no $x$.

The algorithm then looks for a covering of $p$ from
$G_{can}$. There are multiple cases based on how the leading term
of $p$ under the segment being considered appears in the
polynomials in $G_{can}$.
\vspace{-2mm}
\begin{itemize}
	\item
	\textbf{Case 1}: $G_{can} = \emptyset$, then $p$ has no
        covering in $A_i$, implying it is essential. The check
        for $f_3$ in Example \ref{Example Illustrative} is such a case.

	\vspace{-2mm}
      \item \textbf{Case 2}: $q$ in $G_{can}$ such that the
        coefficient of the leading term $t$ of $p$ is determined to be non-zero in
        $A_i$:  $q$ then covers $p$ under $A_i$ iff either it has
        no higher term (In Example \ref{Essentiality Check},
        $f_2$ covers $f_5$ under $A_4$ is such a case),  or the coefficient of all
        of its higher
        terms is 0 (in the above example,  $f_1$ covers $f_4$ under
        $A_4$ is such a case). 
        Otherwise, $q$ cannot cover $p$ if there is a
        higher term with non-zero coefficient, or $q$ can partially cover $p$ if a higher term has a coefficient that is not determined
        to be 0 (as $f_3$ partially covers $f_1$ under $A_1$).
	\item
	\vspace{-2mm}
	\textbf{Case 3}: $q$ in $G_{can}$ such that
        the coefficient of the leading term $t$ of $p$ is not
        determined in $A_i$ to be 0: $q$
        can only partially cover $p$ under $A_i$. $A_i$ is
        partitioned into two segments, and the check is continued
        in each of them. The check of $f_2$ in Example \ref{Example Illustrative} and Example in Section \ref{Section Branch
          Partition} are such cases.
\end{itemize}

\vspace{-2mm}
\subsection{Update CGS}
\label{Section Update CGS}

\ignore{In Example \ref{Example Essential Check}, only $f_1$ is essential
w.r.t.\ $\mathcal{G}$ and $f_1 < f_2 < f_3 < f_4 < f_5$. So
without updating $CGS$, the resulting basis is $S = \{f_1\}$,
which is obviously not even a CGB of $I$. The reason is that
although $f_3$ is non-essential w.r.t.\ $\mathcal{G}$, its only
covering under $A_2$ is $\{f_5\}$. So the removal of $f_5$ makes
$f_3$ essential w.r.t.\ the new CGB $\mathcal{G} - \{f_5\}$. This
is also the case for $f_2$. In general,}

If a polynomial $p$ is found to be not essential, it is discarded
not only from the CGB but it must be replaced by the polynomials
from $CGB$ covering it in the associated $CGS$;
\textit{UpdateCGS} procedure accomplishes that. In every branch
of the CGS in which $p$ appears, $p$ is replaced by the
polynomials covering it for that branch.
As a result, the updated $CGS$
remains to be a CGS of $I$, with the union of polynomials in
$G_i$'s being $\mathcal{G} - \{p\}$, the new CGB.

\ignore{\vspace{-3mm}
\begin{proposition}
\label{Prop Nonessential to Essential}
Given a CGB $\mathcal{G}$ of $I$, let $p$ and $q$ be two
non-essential polynomials w.r.t.\ $\mathcal{G}$. If $q$ is the
only polynomial in $\mathcal{G} - \{p\}$ covering $p$ under some
specialization, then $q$ is essential w.r.t.\ $\mathcal{G}' =
\mathcal{G} - \{p\}$.
\end{proposition}
}

In Example \ref{Example Illustrative}, when removing $f_5$,
$G_4$ is changed to $\{f_4, f_2\}$. Then $f_4$ is still non-essential w.r.t.\
$\mathcal{G} - \{f_5\}$. So remove it, and update $G_4$ 
to be $\{f_1, f_2\}$. After then, all the remaining
polynomials are essential, and $\mathcal{M} = \{f_1, f_2, f_3\}$
is a minimal CGB of $I$.

Sometimes, a branch in $CGS$ may need to be split if $p$ is
covered by two (or more) different polynomials for two (or more)
different parts of the segment. 
%Another issue when updating $CGS$ may happen when a non-essential
%polynomial $p$ has a covering described in Section \ref{Section
%  Branch Partition} over some branch. 
Consider the example in
Section \ref{Section Branch Partition}: in Branch $(A_j =
(\{u^2-v^2\}, \{u\}), G_j = \{p = (u^2-v^2)y+ux\})$, $Q = \{q_1 =
(u+v)x, q_2 = (u-v)x\}$ covers $p$. In case $p$ is
non-essential w.r.t.\ the CGB $\mathcal{G}$ of that ideal, then
Branch $(A_j, G_j)$ must be partitioned into two new branches:
$(A_{j0} = (\{u+v\}, \{u\}), G_{j0} = \{q_2\})$ and $(A_{j1} =
(\{u-v\}, \{u(u+v)\}), G_{j1} = \{q_1\})$. which replace
in the updated CGS, the original branch $(A_j, G_j)$.

\vspace{-2mm}
\subsection{Correctness}
\label{Section Correctness}

\vspace{-4mm}
\begin{proposition}
\label{Prop Correct MCGBMain}
The $\MCGBMain$ algorithm terminates and computes a minimal CGB of the given ideal $I$ w.r.t.\ the given term ordering $>$. 
\end{proposition}
\vspace{-3mm}
\noindent 
\ignore{{\bf Proof Sketch:} Termination of the algorithm
follows the termination of the essentiality check for each
polynomial. That is so because the size of $G_{can}$ decreases or $A_i$
becomes strictly smaller, thus the property of terminating of
\textit{\CEssentialInBranch} follows from K\"{o}nig's Lemma.
%%DK: why do we need Koeing's lemma.
For correctness, the reader should notice the following
invariants of the algorithm: after the essentiality check of
every polynomial, (i) the resulting CGS is still a faithful CGS of the input
ideal with every branch being a reduced \gr basis, (ii) all
branches in the CGS are disjoint, and (iii) the union of the
polynomials in all branches is the CGB of the input ideal. And, only inessential
polynomials are removed from the input CGB.\\
}

\vspace{-2mm}
It should be obvious from the algorithm description that its output is always a subset of the input CGB $\mathcal{G}$.
As stated earlier, the order in which polynomials are checked for
essentiality affects the output of the algorithm. In Example \ref{Example Illustrative}, there are $4$ different MCGBs which are subsets of $\mathcal{G}$:
\vspace{-1mm}
\begin{align*}
	\mathcal{M}_1 &= \{f_1, f_2, f_3\},\\
	\mathcal{M}_2 & = \{f_1, f_3, f_4\},\\
	\mathcal{M}_3 &= \{f_1, f_2, f_5\},\\
	\mathcal{M}_4 &= \{ f_1, f_4, f_5 \}.
\end{align*}

\vspace{-1mm}
It is possible to generate all of these minimal CGBs by changing
the order in which the essentiality check is performed.

Various MCGBs are comparable if the ordering $>$ on polynomials
is extended to sets of polynomials in the usual way.

%DK: fix the set ordering
\ignore{\vspace{-3mm}
\begin{define}
Let $\mathcal{G}_1$ and $\mathcal{G}_2$ be two CGBs of an ideal $I \subseteq K[U][X]$ w.r.t.\ $>$. Let $Monic(\mathcal{G}_1) = \{f_1, f_2, \dots, f_m\}$ and $Monic(\mathcal{G}_2) = \{g_1, g_2, .\dots, g_n\}$, where $Monic$ is to make the leading coefficients of polynomials in the basis monic in $K[U]$, $f_1 > f_2 > \cdots > f_m$ and $g_1 > g_2 > \cdots > g_n$. Then $\mathcal{G}_1 < \mathcal{G}_2$ iff: 
\begin{itemize}
\vspace{-1mm}
\item[(i)]
$\mathcal{G}_1 = \emptyset$ while $|\mathcal{G}_2| > 0$, or
\vspace{-1mm}
\item[(ii)]
$f_1 < g_1$, or
\vspace{-1mm}
\item[(iii)]
$f_1 = g_1$ and $\mathcal{G}_1 - \{f_1\} < \mathcal{G}_2 - \{g_1\}$.
\end{itemize}
\end{define}
}

Using the desecnding order on polynomials in a CGB 
$\mathcal{G}$ 
for the essentiality checking (Line 4 in
\textit{\MCGBMain}), the least MCGB that is a subset of
$\mathcal{G}$ under the set ordering w.r.t.\ $>$ can be generated.

\vspace{-4mm}
\begin{proposition}
\label{Prop Least Subset MCGB}
Given a CGB $\mathcal{G}$ of an ideal $I$ w.r.t.\ $>$ computed by
the KSW algorithm, then \textit{\MCGBMain} algorithm computes the
least MCGB among all MCGBs which are subsets of $\mathcal{G}$
under the set ordering w.r.t.\ $>$.
\end{proposition}

\vspace{-2mm}
For Example  \ref{Example Illustrative}, the algorithm indeed
computes the smallest MCGB $\mathcal{M}_1$.

\subsection{Simplification: Generating a different MCGB from an MCGB}
\label{Section Simplification}

%%DK: although there are references to MCGBsimpl, but there is no
%%algorithm in the paper.

By Proposition \ref{Prop Least Subset MCGB} above,
\textit{\MCGBMain} algorithm 
only computes an MCGB that is the least among all subsets of
the input  CGB $\mathcal{G}$ that are MCGBs. However, this result
need not be the least one among all MCGBs of the ideal $I$ as
illustrated below.

\vspace{-3mm}
\begin{example}
\label{Example Simplification}
For $I = \langle ux^2-2y+(4u+4v)z, (-2u+2v)x^2-2y+4vz
\rangle \subseteq K[u, v][x, y, z]$ and a lexicographic term
order $>$ with $x > y > z \gg u > v$, the CGS computed by the KSW
algorithm is:
\begin{center}
\begin{supertabular}{|c|c|c|c|}\hline
\ & branch & basis & LT \\\hline
$1$ & $(\ \emptyset,\ \{v(3u-2v)\} \ )$ & $\{f_1, f_2\}$ & $\{y, x^2\}$ \\\hline
$2$ & $(\ \{3u-2v\},\ \{ v \}\ )$ & $\{f_4, f_3\}$ & $\{z, x^2\}$ \\\hline
$3$ & $(\ \{u ,\  v \},\ \{1\}\ )$ & $\{f_2\}$ & $\{y\}$ \\\hline
$4$ & $(\ \{v\},\ \{ u \}\ )$ & $\{f_2, f_4\}$ & $\{y, x^2\}$ \\\hline
\end{supertabular}
\end{center}
and the CGB is:
\vspace{-2mm}
\begin{align*}
	\mathcal{G} = \{ & f_1 = (u-\frac{2}{3}v)y+(-\frac{4}{3}u^2-\frac{2}{3}uv+\frac{4}{3}v^2)z,\\
	& f_2 = vx^2-3y+(4u+6v)z,\\
	& f_3 = (u-\frac{10}{13}v)x^2 + \frac{4}{13}y + \frac{12u-8v}{13}z,\\
	& f_4 = (u - \frac{2}{3}v)x^2 + \frac{4}{3}uz \}.
\end{align*}
\end{example}

The output of \textit{\MCGBMain} is $\mathcal{M}_1 = \{f_1, f_2,
f_3\}$. However, even a smaller MCGB can be obtained from 
$\mathcal{M}_1$ by simplifying $f_3$ further. 

Simplification for parametric polynomials can be extremely tricky
however; particularly, replacing a parametric polynomial in a CGB
by another parametric polynomial obtained in general after simplifying using
the CGB need not preserve CGBness.
%As the simplification in \gr bases of polynomial rings,
%analogously, it is also used in $\mathcal{M}$ to achieve a
%smaller MCGB, which is not a subset of $\mathcal{G}$. 
As illustrated earlier, trivial simplification of a polynomial $p$ by another
polynomial $q$  to 0 when $p = a * q$ and $a$ is a polynomial in
parameters, leading to discarding $p$ preserves CGBness (see Example \ref{Example
  Parametric Multiple} as well as Proposition3.5).
The polynomial $-h$, which is a multiple of
$h \in \mathcal{G}$ in the example from Weispfenning discussed in
the Introduction, is also such an example.
However, $h$ can also be simplified to $0$ by $\{f, g\}$, but
removing $h$ does not preserve CGBness since the resulting set
$\{f, g\}$ is not CGB of $I$ anymore. 
\ignore{This is because a CGB of
ideal $I \subseteq K[U][X]$ needs not generate $I$, and the
definition of a CGB is dictated by all possible specializations
of the given ideal, i.e.\ $\sigma(I)$ for $\forall \sigma$,
instead of merely $I$.
%%DK: is the above true?
}

An obvious way to ensure that replacing a polynomial by its
simplified form preserves CGBness is to check that the simplified
form covers the original polynomial along with other polynomials
in the CGB  as defined in the previous section.
That means essentiality check must be performed after simplification
to maintain the correctness. 

We have extended 
 \textit{\MCGBMain} to apply simplification on CGBs
and the result is an extended
MCGB algorithm called \textit{\MCGBSimpl}. The extended algorithm is
conservative in the following sense:  when $p$ is checked to be essential, it is simplified by
$\mathcal{M} - \{p\}$ to the normal form $\widetilde{p}$. If
$\widetilde{p}$ is a non-zero polynomial different from $p$, then
check whether $\widetilde{p}$ with other polynomials in 
$\mathcal{M} - \{p\}$ can cover $p$, in which case
substitute it for $p$ in $\mathcal{M}$ and $CGS$; 
otherwise, keep $p$ in $\mathcal{M}$.

To illustrate the extended algoritm on Example \ref{Example Simplification}
$\mathcal{M} := \mathcal{M} - \{f_4\}$ since $f_4$ is non-essential;
$G_2 = \{f_1, f_3\}$ and $G_4 = \{f_2, f_3\}$. $f_3$ is essential. Instead of keeping it in
$\mathcal{M}$, the algorithm reduces $f_3$ by $\mathcal{M} -
\{f_3\}$ to the normal form $ g = ux^2-2y+(4u+4v)z$. 
The essentiality check on $f_3$ w.r.t.\ $\{f_1, f_2, g\}$
concludes that $f_3$ is non-essential. So
$f_3$ is replaced by $g$ in $M$, and the $CGS$ is updated by setting $G_2 =
\{f_1, f_2\}$ and $G_4 = \{f_2, g\}$.
Both $f_1$ and $f_2$ are essential, and already in their
normal form modulo the other polynomials in
$\mathcal{M}$. The MCGB computed by \textit{\MCGBSimpl} algorithm is
$\{f_1, f_2, g\}$, which is
smaller than $\mathcal{M}_1$ and not a subset of $\mathcal{G}$.

As the reader would have noticed that the extended algorithm is extremely
conservative in applying simplification: if a polynomial $f$ is
essential and its normal form is $g$, $f$ is kept in
$\mathcal{G}$ if the substituted basis fails to be a
CGB. However, there may exist some intermediate form $f'$ along
the reduction chain $f \rightarrow g$, such that $f > f' > g$ and
$(\mathcal{G} - \{f\}) \cup \{f'\}$ remains being a CGB of the
same ideal. So a reasonable modification is to repeatedly perform  one-step
simplification associated with the essentiality check, instead of
checking the normal form only. One step simplification has not
yet been integrated into our implementation, but we plan to do so
in the near future and compare its performance with multi-step
simplification with the goal of generating more and smaller
minimal CGBs.

\vspace{-2mm}
\subsection{Choosing among different Minimal
Dickson Basis in the KSW Algorithm}

An indepth investigation of the KSW algorithm \cite{KSW2} reveals
that for each branch $A_j = (E_j, N_j)$, the corresponding \gr
basis $G_j$ is a minimal Dickson basis (MDB) of the $RGB$ of $F +
N_j$, where $F$ is the given basis. There can in general be many minimal
Dickson bases of the same $RGB$, and the original KSW algorithm
chooses the one with minimal non-zero parts in $A_j$ in its
attempt to compute the least GB (after specialization) for every branch. However,
this choice sometimes results in larger faithful polynomials in
CGS and hence CGB,  producing a larger CGB $\mathcal{G}$ and
thus a larger MCGB as illustrated below.

Consider $F = \{ax^2y+a^2x^2-3a,\ 4ab^2y^2+4b^3-4\} \subseteq K[a, b][x, y]$ with a graded lexicographic term order such that $x > y \gg a > b$. By the original KSW algorithm, we have
\begin{center}
\begin{tabular}{|c|c|c|c|}\hline
\ & branch & basis & LT \\\hline
$1$ & $(\ \emptyset,\ \{ab(a^3b^2+b^3-1)\}\ )$ & $\{f_1, f_2\}$ & $\{y^2, x^2\}$ \\\hline
$2$ & $(\ \{a^3b^2+b^3-1\},\ \{a\}\ )$ & $\{f_3, f_4\}$ & $\{y, x^2\}$ \\\hline
$3$ & $(\ \{a, b^3-1\},\ \{1\}\ )$ & $\emptyset$ & $\emptyset$ \\\hline
$4$ & $(\ \{a\},\ \{b^3-1\}\ )$ & $\{f_1\}$ & $\{1\}$ \\\hline
$5$ & $(\ \{b\},\ \{a\}\ )$ & $\{f_1\}$ & $\{1\}$ \\\hline
\end{tabular}
\end{center}
CGB is
\vspace{-2mm}
\allowdisplaybreaks
\begin{align*}
\mathcal{G} = \{ & f_1 = ab^2y^2+b^3-1,\\
& f_2 = (a^3b^2+b^3-1)x^2+3ab^2y-3a^2b^2,\\
& f_3 = (a^6b^2+2a^3b^3-a^3+b^4-b)x^2\\
& \quad \quad \quad +(3a^4b^2+3ab^3)y-(3a^5b^2+3a^2b^3),\\
& f_4 = (a^6b^2+2a^3b^3-a^3+b^4-b)x^4\\
& \quad \quad \quad -(6a^5b^2+6a^2b^3)x^2+9a^4b^2+9ab^3 \}.
\end{align*}
And the MCGB computed by \textit{\MCGBMain} is $\mathcal{M} = \{f_1, f_2, f_4\}$.

A smaller CGB can be generated however by choosing a minimal
Dickson basis with the
least faithful forms as the \gr basis for each branch. For
the above example, in branch 2, $G_2 = \{f_5, f_6\}$,
can be picked where
\begin{align*}
	f_5 = & (a^5b^2+a^2b^3-a^2)x^2+3a^3b^2y-3a^4b^2,\\
	f_6 = & (a^4b^2+ab^3-a)x^4-6a^3b^2x^2+9a^2b^2,
\end{align*}
with other branches unchanged. So the new CGB $\mathcal{G}' =
\{f_1, f_2, f_5, f_6\}$
resulting in the MCGB $\mathcal{M}' = \{f_1, f_2, f_6\}$ ,
smaller than $\mathcal{M}$ above since $f_1 < f_2 < f_6 < f_4$.

We have implemented this modification to the KSW algorithm to choose the
least minimal Dickson basis w.r.t.\ the 
faithful version of polynomials instead of just considering the
non-zero parts as in the original KSW algorithm with better
results, leading to smaller minimal CGBs.

\vspace{-3mm}
\section{Experiments}
\label{Section Experiment}

The algorithms discussed in the paper are implemented in
SINGULAR \cite{DGPS}. We tested the implementation on a large
suite of examples
including those from \cite{KSW2}, \cite{mccgs}, \cite{gbcover},
\cite{Montes}, \cite{Sit}, \cite{SS3} and \cite{Weis2}. For
instance, there are
$70$ out of $100$ examples which have non-essential polynomials
in the CGBs generated by the KSW algorithm, which is considered
the best algorithm so far for computing smaller CGSs and CGBs
\cite{Montes}. Below we list some of the results.

\begin{center}
\topcaption{Resulted MCGB and CGS}
\label{Table MCGB and CGS}
\vspace{-2mm}
\begin{supertabular}{|c|c|c|c|c|c|} \hline
Example & \parbox[c]{0.8cm}{|\textit{KSW \\ CGS}|} & \parbox[c]{0.8cm}{|\textit{CGS}|}  & $|\mathcal{G}|$ & $|\mathcal{M}|$ & $\%$ reduced \\\hline
bad test & $6$ & $6$ & $8$ & $6$ & $33\%$  \\\hline
KSW51 & $6$ & $5$ & $7$ & $6$ & $17\%$  \\\hline
higher 1 & $4$ & $4$ & $9$ & $6$ & $50\%$  \\\hline
higher 3 & $6$ & $6$ & $6$ & $4$ & $50\%$ \\\hline
linear & $4$ & $4$ & $4$ & $3$ & $33\%$  \\\hline
montes 3 & $12$ & $10$ & $11$ & $6$ & $83\%$  \\\hline
GBCover & $7$ & $5$ & $12$ & $7$ & $71\%$   \\\hline
SS 1 & $4$ & $4$ & $12$ & $10$  & $20\%$  \\\hline
SS 3 & $19$ & $17$ & $36$ & $27$ & $41\%$  \\\hline
Sit 21 & $5$ & $5$ & $6$ & $3$ & $100\%$  \\\hline
Weispfenning 4 & $4$ & $4$ & $3$ & $2$ & $50\%$  \\\hline
Principal & $6$ & $5$ & $3$ & $1$ & $200\%$ \\\hline
CTD & $5$ & $5$ & $6$ & $4$ & $50\%$  \\\hline
S 10 & $4$ & $4$ & $14$ & $12$ & $17\%$  \\\hline
S 12 & $18$ & $15$ & $15$ & $8$ & $88\%$  \\\hline
S 13 & $11$ & $10$ & $9$ & $6$ & $50\%$  \\\hline
S 16 & $19$ & $19$ & $15$ & $9$ & $67\%$  \\\hline
S 53 & $7$ & $5$ & $13$ & $6$ & $117\%$ \\\hline
Nonlinear 1 & $6$ & $6$ & $9$ & $4$ & $125\%$  \\\hline
\end{supertabular}
\end{center}

In Table \ref{Table MCGB and CGS}, the complexity of an example
is characterized by the size of its CGS (i.e. the number of branches)
and CGB, which are the columns with labels $|KSWCGS|$ and
$|\mathcal{G}|$. The size of an MCGB computed by \textit{\MCGBMain}
algorithm is shown in the column with label $|\mathcal{M}|$, with
the percentage of how many non-essential polynomials are removed
from $\mathcal{G}$ caclulated as $(|\mathcal{G}| - |\mathcal{M}|)
/ |\mathcal{M}|$. The column with label $|CGS|$ shows the size of
CGS reconstructed from $\mathcal{M}$.
As the table illustrates, a minimal CGB can reduce the size of
an input CGB by as much as $100\%$ sometimes.

\vspace{-2mm}
\section{Concluding Remarks}

We have proposed an algorithm for computing a minimal CGB from a
given CGB consisting of faithful polynomials. The concept of an
essential polynomial with respect to a CGB is introduced; the
check for essentiality of a polynomial is performed using the
associated CGS by
identifying whether polynomials in other branches can cover the
given polynomial. Only essential polynomials are kept in a CGB
producing a minimal CGB. These two checks only produce minimal
CGBs which are subsets of an input CGB. The concept of a
simplification of an essential polynomial by other essential
polynomials in a CGB is introduced using which minimal CGBs that
are not necessarily subsets of an input CGB can be generated. The
algorithms have been implemented and their effectiveness is
demonstrated on examples which show that most CGBs produced by
various algorithms reported in the literature including the KSW
algorithm have inessential or redundant polynomials. 

From a minimal CGB generated by the proposed algorithm, an
algorithm to compute a CGS from the output 
minimal CGB has been developed; the output CGS of this algorithm is often simpler and
more compact from the original CGS used to generate the minimal
CGB. The discussion of the algorithm could not be included in the
paper becuase of lack of space.

As stated in the introduction, our ultimate goal is to define the
concept of a canonical CGB of a parametric ideal $I$, consisting
of faithful polynomials and uniquely determined by $I$ and term
order $>$. This implies that a canonical CGB should be minimal,
as otherwise, from a nonminimal canonical CGB, it is possible to
identify its proper subset which is both canonical and even
smaller, which contradicts the uniqueness property. Also, it
should be reduced in some sense, since otherwise a canonical and
smaller CGB can be achieved by replacing some polynomial by its
reduced form.

Given an ideal $I \subseteq K[X]$ and a term order $>$, it is easy
to see that the RGB of $I$ is the least \gr basis under the set
ordering w.r.t.\ $>$. However, an analogous definition of
canonical CGB turns out to be difficult.

One major issue is that specialization is not monotonic in
general. To explain it, let $I = \langle f = uz+x, \text{ }g =
(u+1)y-x \rangle \subseteq K[u][z, y, x]$ with a lexicographic term
order $>$ such that $z > y > x \gg u$. By the KSW algorithm, we have
$CGS$
\vspace{-1.5mm}
\begin{center}
\begin{supertabular}{|c|c|c|c|}\hline
$i$ & branch & basis & $\sigma_i(G_i)$ \\\hline
$1$ & $u \neq 0, u+1 \neq 0$ & $\{f, g\}$ & $\{(u+1)y-x,\text{ } uz+x\}$ \\\hline
$2$ & $u+1 = 0$ & $\{g, h\}$ & $\{x, z\}$ \\\hline
$3$ & $u = 0$ & $\{f, h\}$ & $\{x, y\}$ \\\hline
\end{supertabular}
\end{center}

\vspace{-1.5mm} Its CGB $\mathcal{G} = \{f,\ g,\ h = f + g = uz +
(u+1)y\}$ is not minimal, while $\sigma_1(G_1)$, $\sigma_2(G_2)$
and $\sigma_3(G_3)$ are all reduced. A smaller CGB computed by
the \textit{\MCGBMain} algorithm is $\mathcal{M} = \{f, g\}$, which
is also minimal. However, $G_2 = \{g, f\}$ and $G_3 = \{f, g\}$
now, with $\sigma_2(G_2) = \{x, z - x\}$ and $\sigma_3(G_3) =
\{x, y-x\}$ become larger. Namely, we cannot achieve the
canonical CGB by simply reducing the specialized corresponding
\gr bases.

Another issue is that a CGB of an ideal $I$ has no relationship
with its reduced \gr basis (RGB) w.r.t.\ $K[U, X]$. 
It is possible that the CGB generated by the KSW of  algorithm $I$ is
neither a subset of its RGB nor RGB
is a subset of the CGB. So starting with an RGB to compute a
minimal CGB may not be
helpful, especially when RGB is not a CGB. However, if the RGB of
$I$ is a CGB, it can be shown to be minimal as well as canonical CGB.

\vspace{-2mm}
\section*{Acknowledgement}

A preliminary version of this paper was
delivered as an invited talk at \textit{EACA, 2014} in Barcelona, in
June 2014 \cite{KapurYangEACA14}. The first author would like to thank
Prof.\ Yao Sun for many helpful discussions on this
topic. The help provided by Prof.\ Montes through emails is appreciated.

This research project is partially supported by an NSF
  award DMS-1217054. Some of this work was done during the first
  author's sabbatical at the Institute of Software, the Chinese
  Academy of Sciences in Beijing, China.

\end{document}